# Multiphased alkali halide mixed crystals: The conductivity activation energy.

Vassiliki Katsika-Tsigourakou[*] and Chrysoleon Symeonidis

*Section of Solid State Physics, Department of Physics, National and Kapodistrian University of Athens, Panepistimiopolis, 157 84 Zografos, Greece*

**Abstract**

The preparation of mixed crystals of NaBr and KBr from melt as well as their physical characterization have been reported. Electrical measurements were carried out at various temperatures, which showed that their conductivity increases with the increase in temperature and varies nonlinearly with the bulk composition. We find that for most of the compositions studied, the activation energy deduced from the temperature variation of their conductivity obeys a thermodynamical model that interrelates the defect Gibbs energy with the bulk elastic and expansivity data.

--------------------------------------------

* vkatsik@phys.uoa.gr

## 1. Introduction

During the last decades, it became clear that alkali halides mixed crystals enable important applications in optical, optoelectronic and electronic devices [1-5]. This revived the interest on these materials, which have been the object of several early theoretical studies, e.g., see Refs [6-10], that treated the question whether the properties of a mixed system can be described in terms of well known properties of their (pure) constituents.

In view of the aforementioned revived interest, a considerable number of experimental studies on alkali halides mixed systems have appeared [11-21]. The conductivity $\sigma$ of five mixed systems formed when using NaBr and KCl as the starting materials, was measured at various temperatures T [21]. These measurements were fitted into the well-known [22, 23] equation $\sigma = \sigma_0 \exp(-E/k_B T)$, where $k_B$ is the usual Boltzmann's constant, $\sigma_0$ a constant which depends on the material and E the relevant activation energy. Using these E values, and the mean volume per atom $\Omega$ resulting from the density ($\rho$) measurements of Refs [18, 19], as well as the compressibility ($\kappa$) values measured in Ref. [21], we showed [24] that in four out of five systems, the energy E varies linearly with $B\Omega$ where B is the bulk modulus ($=1/\kappa$). This, as discussed in detail in Ref. [24], conforms with an early thermodynamical model [25-35], termed $cB\Omega$ model, which suggests that the defect Gibbs energy $g^i$ for a certain process i under study (i.e., i=formation, migration, self-diffusion, activation, etc.) is proportional to $B\Omega$ according to the relation

$$g^i = c^i B\Omega \tag{1}$$

where $c^i$ is a dimensionless constant. This model has been found [10, 25-35] to agree with the experimental data for various categories of solids and defect processes. Of particular importance is the case [36] where, upon applying uniaxial stress on ionic crystals, electric signals are produced due to defect formation and migration. This helps [36] towards understanding the experimental fact that transient electric signals are detected [37-47] before the occurrence of major earthquakes.

The aforementioned publication [24] was crossed with the report [19] of measurements in mixed crystals of NaBr and KBr that were also prepared from melt and physically characterized. It is the object of this short paper to investigate whether the aforementioned finding in the mixed crystals of NaBr and KCl between E and $B\Omega$ holds for the newly prepared mixed systems of NaBr and KBr as well.

## 2. The new data and analysis

The following five mixed systems were found [19] by using NaBr and KBr as the starting materials: MC1: $Na_{0.2}K_{0.8}Br$; MC2: $Na_{0.4}K_{0.6}Br$; MC3: $Na_{0.5}K_{0.5}Br$; MC4: $Na_{0.6}K_{0.4}Br$ and MC5: $Na_{0.8}K_{0.2}Br$. Among other studies, dc conductivity and dielectric measurements have been performed [19] with the standard procedure (see Refs [22, 23] for the former and latter measurements, respectively) in all the aforementioned five mixed systems as well as in the two pure (polycrystalline) end members. The corresponding E values deduced from the conductivity measurements (as read from Fig. 9 of Ref. [19]) are given in Table 1.

In order to answer the question under investigation, we need to know the values of B and $\Omega$ for each of the aforementioned crystals. The following procedures have been applied: First, concerning the $\Omega$ values, we followed the procedure explained in our previous publication [24]. As for the B values, we used the compressibility values measured by Padma and Mahadevan [19], which are tabulated in the last but one column of their Table 3, and then applied the relation $B = 1/\kappa$. In this way, the $B\Omega$ value was obtained for each of the aforementioned seven crystals and these results are presented in Table 1.

## 3. Results and Discussion

In Fig. 1 we plot E versus $B\Omega$ for the seven crystals investigated, i.e. the five mixed alkali halides systems MC1 to MC5 and the two end members NaBr and KBr. An inspection of this figure shows that for four out of seven systems investigated, i.e., KBr, MC1, MC2 and MC4, the activation energy E increases with the increase of $B\Omega$.

In addition, the points corresponding to the latter four crystals mentioned above seem to scatter around the straight line that has been drawn in Fig. 1 on the basis of the $cB\Omega$ model as follows: By differentiating Eq. (1) in respect to temperature one finds the activation entropy according to $s^i = -(dg^i/dT)_P$ and therefrom we get the activation enthalpy $h^i$ through the well known relation $h^i = g^i + Ts^i$. This leads to:

$$h^i = c^i\Omega\left\{B - T\beta B - T\left(dB/dT\right)_P\right\} \qquad (2)$$

where $\beta$ is the thermal (volume) expansion coefficient. In addition, the activation volume $v^i$, defined as $v^i = \left(dg^i/dP\right)_T$, is found to be:

$$v^i = c^i \Omega \left[\left(dB/dP\right)_T - 1\right] \quad (3)$$

Thus, combining Eqs (2) and (3), we get:

$$\frac{h^i}{v^i} = \frac{B - T\beta B - T\left(dB/dT\right)_P}{\left(dB/dP\right)_T - 1}$$

which, when considering the definition of the Anderson-Gruneisen parameter $\delta$, i.e.,

$$\delta \equiv -\frac{1}{\beta B}\left(\frac{dB}{dT}\right)_P \quad (4)$$

turns to:

$$h^i = \frac{v^i}{\Omega} \times \frac{(1 - T\beta + T\beta\delta)}{\left(dB/dP\right)_T - 1} B\Omega \quad (5)$$

Considering reasonable values for $\frac{v^i}{\Omega}$ in the extrinsic range [48] of the conductivity curve, i.e. $\frac{v^i}{\Omega} \approx 0.4$, as well as the elastic and expansivity data given in Ref. [49, 50], we find that $h^i$ vs $B\Omega$ may be approximated with a straight line (since the quantities $\delta$ and $(dB/dP)_T$ have small variations among the alkali halides) having a slope of around $8 \times 10^{-2}$ with a plausible experimental error of around 20%. (Note that the activation enthalpy $h^i$ coincides with E for the case of linear conductivity plots). This is the slope of the straight line drawn in Fig. 1.

The reason why the points corresponding to the two crystals MC3 and MC4, deviate markedly (i.e., outside the experimental error mentioned above) from the behaviour predicted by the $cB\Omega$ model, is not clear. This is striking because it has been found [10] that the thermodynamic parameters for Shottky defect formation, as well as for cation vacancy migration (in monocrystalline alkali halides) do obey the $cB\Omega$ model. This indicates that the following might have happened in the polycrystalline samples of MC3 and MC4 measured by Padma and Mahadevan [19]: they presumably contain a considerable number of grain boundaries, which affects [10] considerably the defect migration processes. In order to confirm this possibility, additional experiments are necessary.

## 4. Conclusions

Using experimental data alone, we investigated the activation energy E for the increase of the conductivity with the increase of temperature as a function of $B\Omega$ for the mixed crystals of NaBr and KBr. We find that, in four out of seven systems, E varies linearly with $B\Omega$ which in addition, has a slope equal to that predicted by the $cB\Omega$ model. These results seem to strengthen the validity of those obtained in our independent study for the mixed systems of NaBr and KCl [24].

**FIGURE & FIGURE CAPTION**

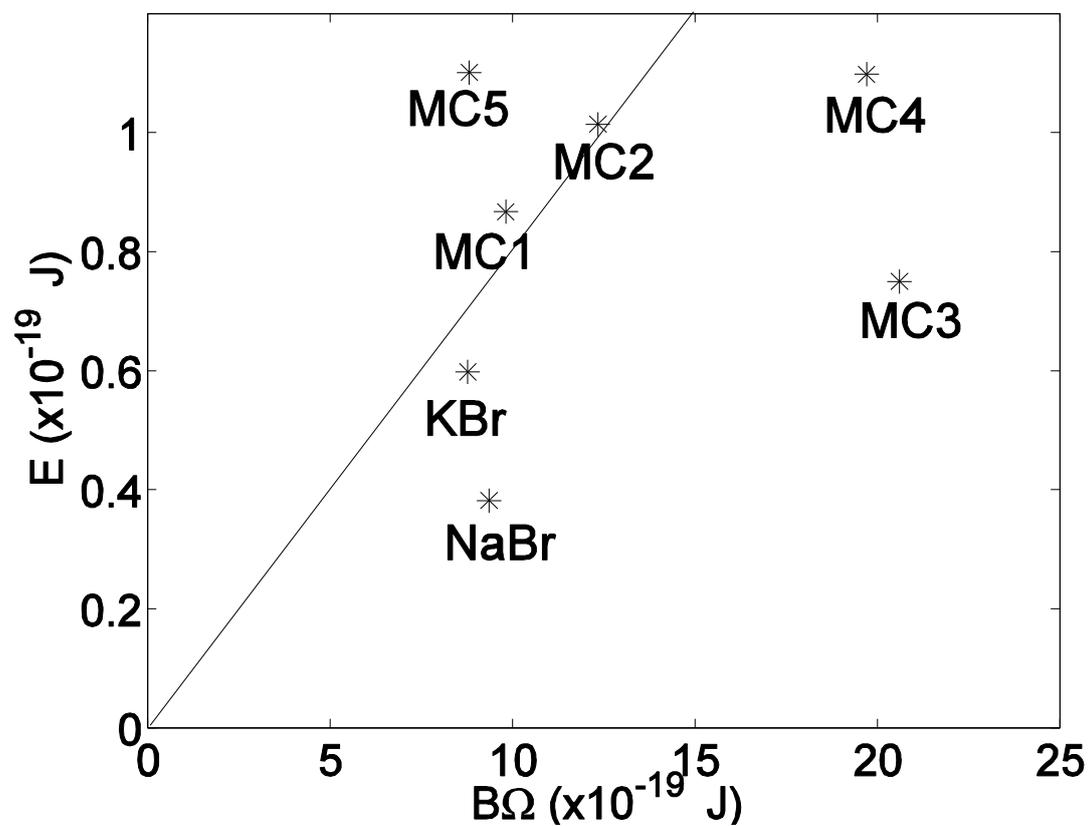

**Fig. 1.** The experimental values of the activation energy E, reported by Padma and Mahadevan [19], versus $B\Omega$. The straight line depicts the prediction of the $cB\Omega$ model (see the text).

**Table 1.**

The values of the activation energy E, the compressibility $\kappa$ and the density $\rho$ reported in Ref. [19] for the five mixed systems of NaBr and KBr and the two end members. The values of $B\Omega$ given in the last column, where $B(=1/k)$ is the bulk modulus and $\Omega$ the mean volume per atom, are treated in the study explained in the text.

| System[a] | E | $\kappa$ | $\rho$ | $B\Omega$ |
|---|---|---|---|---|
| | $10^{-19}$ J | $(10^{-11}\,\text{m}^2/\text{N})$ | (gr/cc) | $10^{-19}$ J |
| NaBr | 0.382 | 5.725 | 3.1882 | 9.359 |
| KBr | 0.598 | 8.499 | 2.6485 | 8.777 |
| MC1: $Na_{0.209}K_{0.791}Br$ | 0.867 | 8.953 | 2.7612 | 9.818 |
| MC2: $Na_{0.362}K_{0.638}Br$ | 1.014 | 8.389 | 2.8437 | 12.345 |
| MC3: $Na_{0.461}K_{0.539}Br$ | 0.750 | 5.755 | 2.8981 | 20.606 |
| MC4: $Na_{0.541}K_{0.459}Br$ | 1.098 | 5.837 | 2.9403 | 19.703 |
| MC5: $Na_{0.827}K_{0.173}Br$ | 1.101 | 7.775 | 3.0946 | 8.818 |

[a] estimated bulk composition in the crystal [19]